# Magnetic and transport studies of the TbAgAl compound at high fields

Ashwin Mohan[1*], S. Radha[2]
[1] Department of Physics, Institute of Chemical Technology, Mumbai-19
[2] Dept of Physics, University of Mumbai, Vidyanagari, Mumbai-98

*work carried out at Department of Condensed Matter Physics, TIFR, Mumbai-5 in 2010

## Abstract:

In order to further investigate the magnetic state of the RAgAl series, the magnetization measurements on the TbAgAl compound from this series have been extended to higher fields of 12 Tesla in the temperature range 2K-300K. The electrical resistivity in the temperature range 2-300K has been measured up to fields of 9 Tesla. The field dependence of magnetization at low temperatures suggests an antiferromagnetic state undergoing a metamagnetic transition to a ferromagnetic state above the critical field. The observation of large coercivity (unlike other compounds in the RAgAl series) and non-saturation of magnetization indicates a disordered magnetic state having both ferromagnetic and antiferromagnetic exchange interaction. The presence of competing interactions leading to a disordered state is also supported by transport measurements and is attributed presumably to the layered structure of the compound.

# Introduction:

The RAgAl (R = rare earth) compounds crystallizing in the orthorhombic CeCu2 type structure, have been studied for the reported magnetocaloric effect [1,2]. The magnetization, AC susceptibility and heat capacity measurements have suggested a short-range order with the spin-glass like state [1,3] or a complex Griffiths phase [2]. The magnetic studies of the compound are further extended to higher fields of 12 T in the temperature range 2-300K.

# Experimental details:

The TbAgAl compound was prepared by arc melting stoichiometric amounts of high purity constituent elements. It was melted three to four times and then annealed in sealed quartz ampoules at 900 C for 1 week. It was structurally characterized by x-ray powder diffraction and the $CeCu_2$ structure was confirmed. Magnetization measurements in the temperature range 2K-300K and fields up to 12 T were made on the Vibrating Sample Magnetometer while the SQUID magnetometer (Quantum Design MPMS5) was also used for temperature dependence of magnetization. Transport measurements in the temperature range 2-300 K and fields of 9 T were made on the Physical Property Measurement System (Quantum Design PPMS).

# Results and Discussions:

The temperature dependence of magnetization is shown in fig 1(a-b). The low field data at 50 Oe shows a peak in magnetization at 62K indicating the onset of magnetic ordering. A strong irreversibility in magnetization between zero field cooled (ZFC) and field cooled (FC) state is observed below this peak temperature $T_c$. The peak temperature in the ZFC data shifts significantly to lower temperatures with increasing magnetic field, reaching a value of 15K in a field of 2T. The variation of this peak temperature with field is shown in the inset of figure 1b.

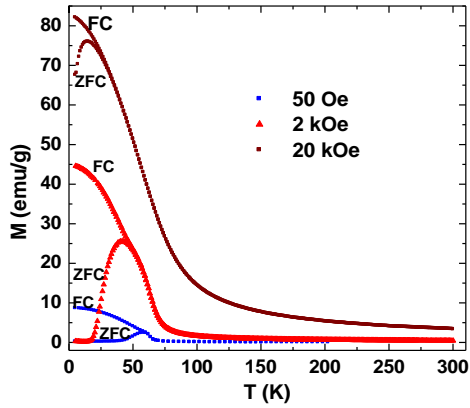 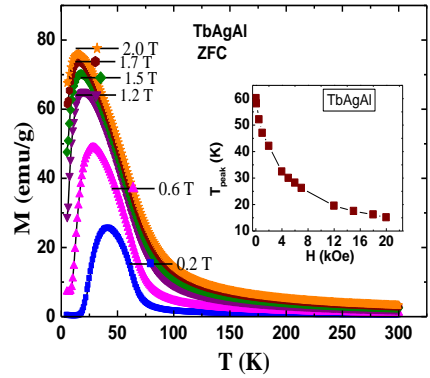

| Fig 1a Temp (T) dependence of M in fields 50 Oe to 20 kOe in ZFC and FC | Fig 1b M vs T in ZFC state in varying fields. Inset shows field variation of peak temp Tc |

The variation of magnetization (M) with external field (H) is shown in figures 2(a-b). At 2K, a sharp step-like metamagnetic transition is observed at a field of 1.6T. Further, the magnetization reversal at Coercive field ($H_c$) up to 3 K, is even more step-like. At these temperatures, the virgin curve is outside the M-H loop as in Figure 2a. Above 4 K, the field induced metamagnetic transition is sharp but gradual. Further the virgin curve now lies within the M-H loop, as is the case with normal ferromagnets. At all temperatures, the magnetization does not saturate even in the highest measurement field of 12T (carried out at few temperatures). A relatively large coercivity (~1.4T at 2K and 1T at 8K) is observed at low temperatures and it decreases nearly exponentially with increasing temperature. This variation is shown in the inset of figure 2b.

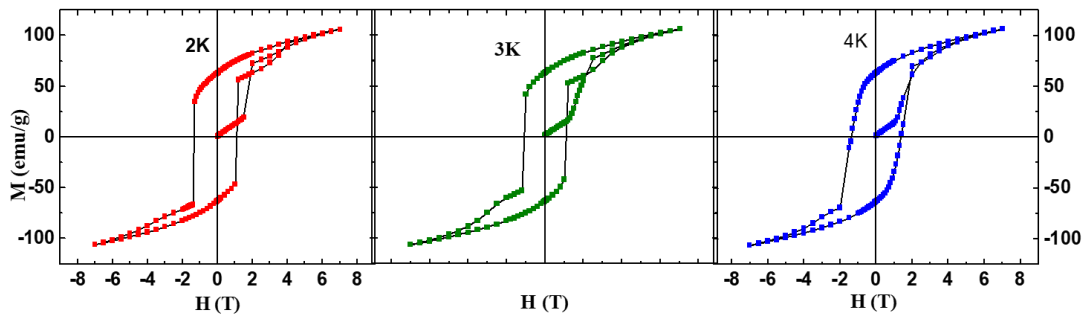

**Fig 2a** Field dependence of magnetization at 2K, 3K and 4K

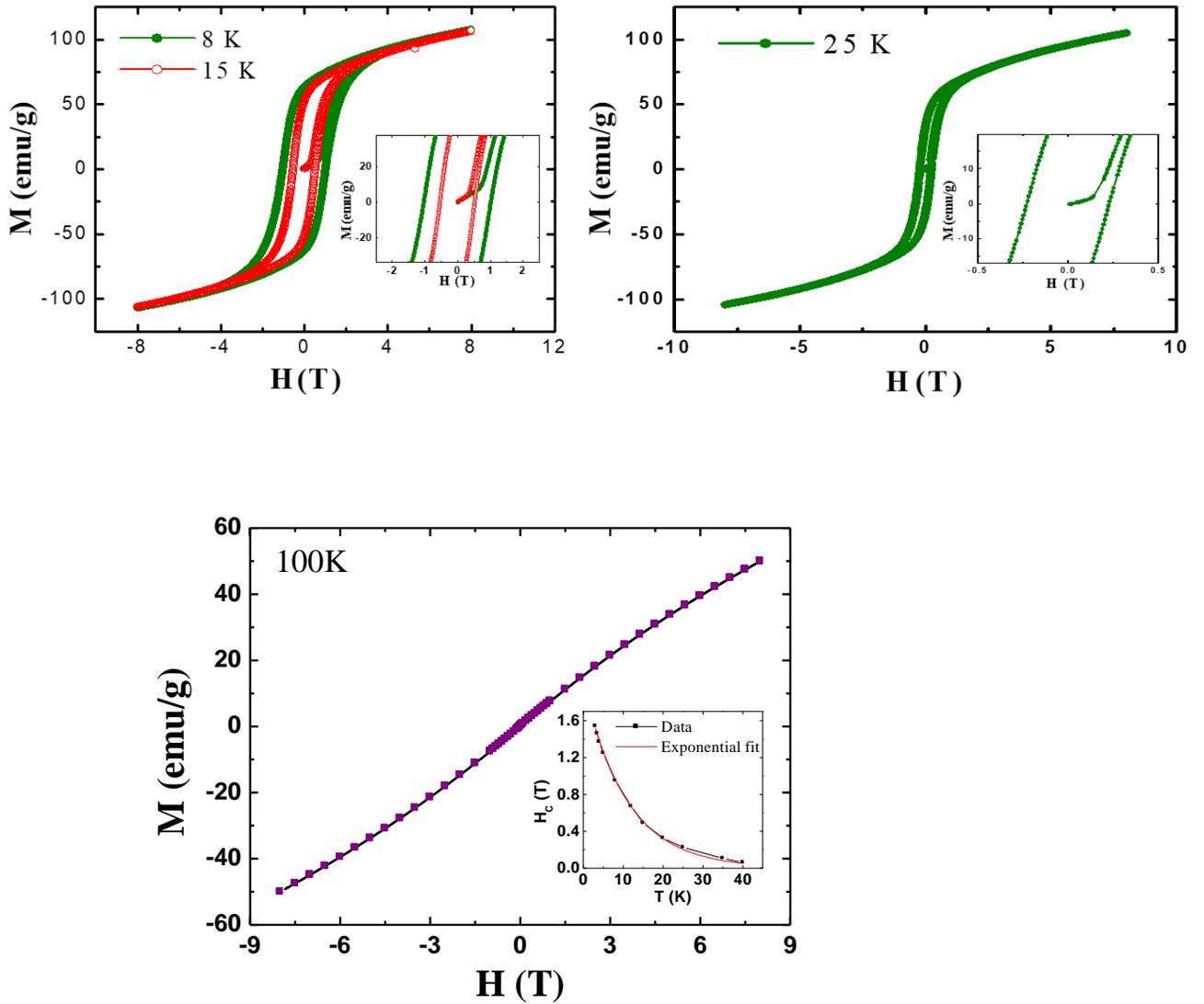

**Fig 2b** Field dependence of magnetization between 8K and 100K. Inset in last plot shows variation of coercivity with temperature with an exponential fit.

The temperature variation of Resistance (R) in zero field and in 1.5 T is shown in Figure 3a. A knee-like anomaly is observed around $T_c$ and a linear variation of R on T is observed at temperatures above 100K. The field dependence of magnetoresistance ($\Delta\rho/\rho$) at 20K and 70K is shown in the inset of Figure 3a. Figure 3b shows the field dependence of resistivity ($\rho$) in fields up to 90 kOe and temperature 2K-70K. The magnetoresistance is found to be negative and small: it is around 1.6% at 20K at the highest measured fields of 9 T.

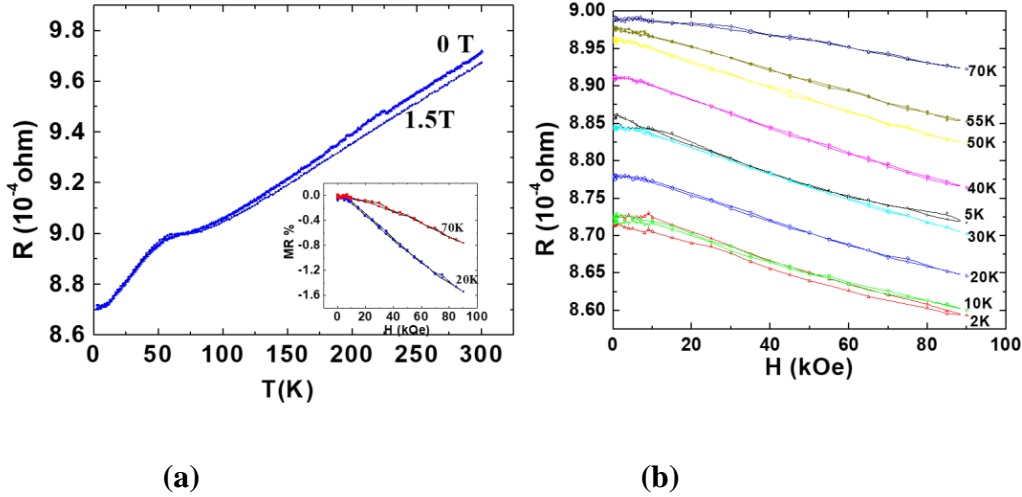

(a)                  (b)

**Fig 3: (a)** Temperature dependence of Resistance in zero field and 1.5T. Inset shows Field variation of MR(%) in 20K and 70K. **(b)** Field dependence of Resistance in temp range 2-70K.

The temperature dependence of magnetization suggests that a dominant ferromagnetic transition occurs at around 60 K. With decreasing temperature, the antiferromagnetic exchange interaction grows resulting in strong field dependence of the peak temperature $T_c$. The above result is supported by the observation of a metamagnetic transition at low temperatures in the field dependence of magnetization. The earlier study of relaxation dynamics of magnetization [2] has suggested the existence of a spin glass like state as a result of competing ferro- and antiferro- magnetic interactions. It is possible that the low field state is frustrated due to the competing magnetic exchange interactions leading to a spin glass behavior. In higher fields, ferromagnetic exchange interaction dominates, as is suggested by the presence of high coercivity that is observed in the magnetic hysteresis loops, particularly at low temperatures. However, the non-saturation of magnetization even in a field of 12 Tesla indicates presence of antiferromagnetic exchange interaction in the compound.

The field induced step-like transition has been observed in some manganites and rare earth intermetallics with non-linear spin structures [6,7], as a result of martensitic character of antiferro- to ferro-magnetic transition. It is difficult to envisage a similar situation in *TbAgAl*.
.

## Conclusion:

The results of high field measurements on the TbAgAl compound suggest the growth of ferromagnetic exchange interaction that leads to observation of large coercivity and strong field dependence of the peak temperature in magnetization. The observed metamagnetic transition at low temperatures from an antiferromagnetic to a ferromagnetic state, fortifies the presence of competing interactions, and its origin is presumably attributed to the layered-structure of the compound [4,5]. The existence of a spin-glass like state is possible, as a result of the competing antiferro- and ferro-magnetic exchange interaction present in the compound at low fields. The measurements of electrical resistivity and heat capacity [2] are also in tune with the spin glass like state in low fields. Further systematic magnetic and specific heat studies of compounds with varying rare earth substitutions and supporting neutron diffraction data are essential to establish the nature of magnetic ordering in these compounds.

**Acknowledgments:** The authors express their sincere thanks to the continuous support received from Prof A.K.Nigam and staff of Department of Condensed Matter Physics, TIFR, Mumbai for this collaborative work carried out during 2009-10, a part of which was presented at the International Conference on Strongly Correlated Electron Systems (SCES 2011) at Cambridge, UK. The help received from Mr Shaikh Arif, AIKTC, Panvel for help in curating the data is duly acknowledged.